\documentclass{article}

\usepackage{color}
\usepackage{amsmath, amssymb, amsthm}

\newcommand{\RM}{\mathbb{R}}
\newcommand{\ZM}{\mathbb{Z}}

\newtheorem{theorem}{Theorem}
\newtheorem{lemma}{Lemma} 
\newtheorem{prop}{Proposition} 
\newtheorem{cor}{Corollary}

\newtheorem*{proof*}{Proof}

\newcommand{\xvec}{\ensuremath{\boldsymbol{x}}}

\newcommand{\evec}{\ensuremath{\boldsymbol{e}}}

\newcommand{\kvec}{\ensuremath{\boldsymbol{k}}}
\newcommand{\wvec}{\ensuremath{\boldsymbol{w}}}

\newcommand{\disf}{\displaystyle\frac}

\begin{document}

\title{{\bf Walk/Zeta Correspondence \\ 
for quantum and correlated random walks}
\vspace{15mm}}

\author{
Norio KONNO \\
Department of Applied Mathematics, Faculty of Engineering \\ 
Yokohama National University \\
Hodogaya, Yokohama, 240-8501, JAPAN \\
e-mail: konno-norio-bt@ynu.ac.jp \\
\\
Shunya TAMURA \\
Graduate School of Science and Engineering \\
Yokohama National University \\ 
Hodogaya, Yokohama, 240-8501, JAPAN \\ 
e-mail: tamura-shunya-kj@ynu.jp 
}

\date{\empty }

\maketitle

\vspace{50mm}


\vspace{20mm}


%









\clearpage

\begin{abstract}
In this paper, following the recent paper on Walk/Zeta Correspondence by the first author and his coworkers, we compute the zeta function for the three- and four-state quantum walk and correlated random walk, and the multi-state random walk on the one-dimensional torus by using the Fourier analysis. We deal with also the four-state quantum walk and correlated random walk on the two-dimensional torus. In addition, we introduce a new class of models determined by the generalized Grover matrix bridging the gap between the Grover matrix and the positive-support of the Grover matrix. Finally, we give a generalized version of the Konno-Sato theorem for the new class. As a corollary, we calculate the zeta function for the generalized Grover matrix on the $d$-dimensional torus. 
\end{abstract}

\vspace{10mm}

\begin{small}
\end{small}

\vspace{10mm}

\section{Introduction \label{sec01}}
The quantum walk (QW) is a quantum counterpart of the correlated random walk (CRW). Note that the random walk (RW) is a special model of the CRW. The Grover walk is one of the most well-investigated models in the study of the QW. The zeta function for the Grover walk can be obtained by the so-called Konno-Sato theorem given in \cite{KS}. Recently, the first author and his coworkers gave explicit formulas for the generalized zeta function and the generalized Ihara zeta function corresponding to the flip-flop type Grover walk and its positive-support version respectively, on  a class of graphs by using the Konno-Sato theorem in \cite{K1}. This relation is called ``Grover/Zeta Correspondence''. In the subsequent paper \cite{K2}, they obtained the zeta function for the wide class of walks including RW, CRW, QW, and open quantum random walk (OQRW) on the torus. The relation is called ``Walk/Zeta Correspondence''. Moreover, their paper \cite{K4} presented the characteristic polynomial of the transition matrix related to the vertex-face walk on the two-dimensional torus. In these papers \cite{K1, K2, K4}, they dealt with one-particle models including RW, CRW, QW, OQRW. On the other hand, they investigated multi-particle models with probabilistic or quantum interactions, called the interacting particle systems (IPS) in \cite{K3}. The relation between this model and the zeta function is called ``IPS/Zeta Correspondence''. In addition, the model studied in \cite{K1, K2, K3, K4} is the discrete-time model, while they studied corresponding continuous-time model (CTM) in \cite{K5}, called ``CTM/Zeta Correspondence". 

In this paper, we extend the models studied in \cite{K2} and calculate the zeta function for the extended classes of models on the one- and two-dimensional torus via Walk/Zeta Correspondence. Furthermore, we introduce a new class of models determined by the generalized Grover matrix bridging the gap between the Grover matrix and the positive-support of the Grover matrix. Finally, we present a generalized version of the Konno-Sato theorem for the new class. As a corollary, we compute the zeta function for the generalized Grover matrix on the $d$-dimensional torus. 

The rest of this paper is organized as follows. In Section \ref{sec02}, we review Walk/Zeta Correspondence on the torus investigated in \cite{K2}. Sections \ref{sec03} and \ref{sec04} are devoted to the three- and four-state QW and CRW on the one-dimensional torus. In Section \ref{sec05}, we treat the multi-state RW on the one-dimensional torus. Section \ref{sec06} deals with the four-state QW and CRW on the two-dimensional torus. Section \ref{sec07} introduces a new class of models determined by the generalized Grover matrix. In Section \ref{sec08}, a generalized version of the Konno-Sato theorem for the new class is given. By using this theorem, we calculate the zeta function for the generalized Grover matrix on the $d$-dimensional torus. Furthermore, we mention a relation between Grover/Zeta and Walk/Zeta Correspondences. Section \ref{sec09} summarizes our results.

\section{Walk/Zeta Correspondence \label{sec02}}
First we introduce the following notation: $\mathbb{Z}$ is the set of integers, $\mathbb{Z}_{\ge}$ is the set of non-negative integers, $\mathbb{Z}_{>}$ is the set of positive integers, $\mathbb{R}$ is the set of real numbers, and $\mathbb{C}$ is the set of complex numbers. Moreover, $T^d_N$ denotes the {\em $d$-dimensional torus} with $N^d$ vertices, where $d, \ N \in \mathbb{Z}_{>}$. Remark that $T^d_N = (\mathbb{Z} \ \mbox{mod}\ N)^{d}$.

Following \cite{K2} in which Walk/Zeta Correspondence on $T^d_N$ was investigated, we will review our setting for $2d$-state discrete time walk with a nearest-neighbor jump on $T^d_N$. The present paper will deal with the general $d_c$-state walk on $T^d_N$ with $d_c \in \ZM_{>}$. However, it is easy to understand our setting for the case of $2d$-state walk with the nearest-neighbor jump, so we treat this type of model.

The discrete time walk is defined by using a {\em shift operator} and a {\em coin matrix} which will be mentioned below. Let $f : T^d_N \longrightarrow \mathbb{C}^{2d}$. For $j = 1,2,\ldots,d$ and $\xvec \in T^d_N$, the shift operator $\tau_j$ is defined by 
\begin{align*}
(\tau_j f)(\xvec) = f(\xvec-\evec_{j}),
\end{align*} 
where $\{ \evec_1,\evec_2,\ldots,\evec_d \}$ denotes the standard basis of $T^d_N$. The {\em coin matrix} $A=[a_{ij}]_{i,j=1,2,\ldots,2d}$ is a $2d \times 2d$ matrix with $a_{ij} \in \mathbb{C}$ for $i,j =1,2,\ldots,2d$. If $a_{ij} \in [0,1]$ and $\sum_{i=1}^{2d} a_{ij} = 1$ for any $j=1,2, \ldots, 2d$, then the walk is a CRW. We should remark that, in particular, when $a_{i1} = a_{i2} = \cdots = a_{i 2d}$ for any $i=1,2, \ldots, 2d$, this CRW becomes a RW. If $A$ is unitary, then the walk is a QW. So our class of walks contains RW, CRW, and QW as special models. 

To describe the evolution of the walk, we decompose the $2d \times 2d$ coin matrix $A$ as
\begin{align*}
A=\sum_{j=1}^{2d} P_{j} A,
\end{align*}
where $P_j$ denotes the orthogonal projection onto the one-dimensional subspace $\mathbb{C}\eta_j$ in $\mathbb{C}^{2d}$. Here $\{\eta_1,\eta_2, \ldots, \eta_{2d}\}$ denotes a standard basis on $\mathbb{C}^{2d}$.

The discrete time walk associated with the coin matrix $A$ on $T^d_N$ is determined by the $2d N^d \times 2d N^d$ matrix
\begin{align}
M_A=\sum_{j=1}^d \Big( P_{2j-1} A \tau_{j}^{-1} + P_{2j} A \tau_{j} \Big).
\label{unitaryop1}
\end{align}
The state at time $n \in \mathbb{Z}_{\ge}$ and location $\xvec \in T^d_N$ can be expressed by a $2d$-dimensional vector:
\begin{align*}
\Psi_{n}(\xvec)=
\begin{bmatrix}
\Psi^{1}_{n}(\xvec) \\ \Psi^{2}_{n}(\xvec) \\ \vdots \\ \Psi^{2d}_{n}(\xvec) 
\end{bmatrix} 
\in \mathbb{C}^{2d}.
\end{align*}

For $\Psi_n : T^d_N \longrightarrow \mathbb{C}^{2d} \ (n \in \mathbb{Z}_{\geq})$, Eq. \eqref{unitaryop1} gives the evolution of the walk as follows.
\begin{align}
\Psi_{n+1}(\xvec) \equiv (M_A \Psi_{n})(\xvec)=\sum_{j=1}^{d}\Big(P_{2j-1}A\Psi_{n}(\xvec+\evec_j)+P_{2j}A\Psi_{n}(\xvec-\evec_j)\Big).
\label{reunitaryop1}
\end{align} 
This equation means that the walker moves at each step one unit to the $- x_j$-axis direction with matrix $P_{2j-1}A$ or one unit to the $x_j$-axis direction with matrix $P_{2j}A$ for $j=1,2, \ldots, d$. Moreover, for $n \in \ZM_{>}$ and $\xvec = (x_1, x_2, \ldots, x_d) \in T^d_N$, the $2d \times 2d$ matrix $\Phi_n (x_1, x_2, \ldots, x_d)$ is given by 
\begin{align*}
\Phi_n (x_1, x_2, \ldots, x_d) = \sum_{\ast} \Xi_n \left(l_1,l_2, \ldots , l_{2d-1}, l_{2d} \right),
\end{align*} 
where the $2d \times 2d$ matrix $\Xi_n \left(l_1,l_2, \ldots , l_{2d-1}, l_{2d} \right)$ is the sum of all possible paths in the trajectory of $l_{2j-1}$ steps $- x_j$-axis direction and  $l_{2j}$ steps $x_j$-axis direction and $\sum_{\ast}$ is the summation over $\left(l_1,l_2, \ldots , l_{2d-1}, l_{2d} \right) \in (\ZM_{\ge})^{2d}$ satisfying 
\begin{align*}
l_1 + l_2 + \cdots + l_{2d-1} + l_{2d} = n, \qquad x_j = - l_{2j-1} + l_{2j} \quad (j=1,2, \ldots, d).
\end{align*} 
Here we put
\begin{align*}
\Phi_0 (x_1, x_2, \ldots, x_d)
= \left\{ 
\begin{array}{ll}
I_{2d} & \mbox{if $(x_1, x_2, \ldots, x_d) = (0, 0, \ldots, 0)$, } \\
O_{2d} & \mbox{if $(x_1, x_2, \ldots, x_d) \not= (0, 0, \ldots, 0)$},
\end{array}
\right.
\end{align*}
where $I_n$ is the $n \times n$ identity matrix and $O_n$ is the $n \times n$ zero matrix. Then, for the walk starting from $(0,0, \ldots, 0)$, we obtain  
\begin{align*}
\Psi_n (x_1, x_2, \ldots, x_d) = \Phi_n (x_1, x_2, \ldots, x_d) \Psi_0 (0, 0, \ldots, 0) \qquad (n \in \mathbb{Z}_{\ge}).
\end{align*} 
We call $\Phi_n (\xvec) = \Phi_n (x_1, x_2, \ldots, x_d)$ {\em matrix weight} at time $n \in \mathbb{Z}_{\ge}$ and location $\xvec \in T^d_N$ starting from ${\bf 0} = (0,0, \ldots, 0)$. When we consider the walk on not $T^d_N$ but $\ZM^d$, we add the superscript ``$(\infty)$" to the notation like $\Psi^{(\infty)}$ and $\Xi^{(\infty)}$.

This type is {\em moving} shift model called {\em M-type} here. Another type is {\em flip-flop} shift model called {\em F-type} whose coin matrix is given by 
\begin{align}
A^{(f)} = \left( I_{d} \otimes \sigma \right) A,
\label{goho}
\end{align} 
where $\otimes$ is the tensor product and
\begin{align*}
\sigma 
=
\begin{bmatrix}
0 & 1 \\ 
1 & 0 
\end{bmatrix} 
. 
\end{align*}
For example, when $d = 2$ (two-dimensional case), we have
\begin{align*}
I_2 \otimes \sigma = 
\begin{bmatrix}
\sigma & O_2 \\ 
O_2 & \sigma
\end{bmatrix} 
.
\end{align*} 
The F-type model is also important, since it has a central role in the Konno-Sato theorem \cite{KS}. When we distinguish $A$ (M-type) from $A^{(f)}$ (F-type), we write $A$ by $A^{(m)}$.

The measure $\mu_n (\xvec)$ at time $n \in \mathbb{Z}_{\ge}$ and location $\xvec \in T^d_N$ is defined by
\begin{align*}
\mu_n (\xvec) = \| \Psi_n (\xvec) \|_{\mathbb{C}^{2d}}^p = \sum_{j=1}^{2d}|\Psi_n^{j}(\xvec)|^p,
\end{align*} 
where $\|\cdot\|_{\mathbb{C}^{2d}}^p$ denotes the standard $p$-norm on $\mathbb{C}^{2d}$. As for CRW and QW, we take $p=1$ and $p=2$, respectively. Then CRW and QW satisfy 
\begin{align*}
\sum_{\xvec \in T_N^d} \mu_n (\xvec) = \sum_{\xvec \in T_N^d} \mu_0 (\xvec), 
\end{align*} 
for any time $n \in \ZM_{>}$. However, we do not necessarily impose such a condition for the walk we consider here. For example, the two-dimensional positive-support version of the Grover walk (introduced in Section \ref{sec07}) does not satisfy the condition. In this meaning, our walk is a generalized version for the usual walk.

To consider the zeta function, we use the Fourier analysis. To do so, we introduce the following notation: $\mathbb{K}_N = \{ 0,1, \ldots, N-1 \}$ and $\widetilde{\mathbb{K}}_N = \{ 0 ,2 \pi/N, \ldots, 2 \pi (N-1)/N \}$.

For $f : \mathbb{K}_N^d \longrightarrow \mathbb{C}^{2d}$, the Fourier transform of the function $f$, denoted by $\widehat{f}$, is defined by the sum
\begin{align}
\widehat{f}(\kvec) = \frac{1}{N^{d/2}} \sum_{\xvec \in \mathbb{K}_N^d} e^{- 2 \pi i \langle \xvec, \kvec \rangle /N} \ f(\xvec),
\label{yoiko01}
\end{align}
where $\kvec=(k_1,k_2,\ldots,k_{d}) \in \mathbb{K}_N^d$. Here $\langle \xvec,  \kvec \rangle$ is the canonical inner product of $\mathbb{R}^d$, i.e., $\langle \xvec,  \kvec \rangle = \sum_{j=1}^{d} x_j k_j$. Then we see that $\widehat{f} : \mathbb{K}_N^d \longrightarrow \mathbb{C}^{2d}$. Moreover, we should remark that 
\begin{align}
f(\xvec) = \frac{1}{N^{d/2}} \sum_{\kvec \in \mathbb{K}_N^d} e^{2 \pi i \langle \xvec, \kvec \rangle /N} \ \widehat{f}(\kvec),
\label{yoiko02}
\end{align}
where $\xvec =(x_1,x_2,\ldots,x_{d}) \in \mathbb{K}_N^d$. By using 
\begin{align}
\widetilde{k}_j = \frac{2 \pi k_j}{N} \in \widetilde{\mathbb{K}}_N, \quad \widetilde{\kvec}=(\widetilde{k}_1,\widetilde{k}_2,\ldots,\widetilde{k}_{d}) \in \widetilde{\mathbb{K}}_N^d, 
\label{kantan01}
\end{align}
we can rewrite Eqs. \eqref{yoiko01} and \eqref{yoiko02} in the following way: 
\begin{align*}
\widehat{g}(\widetilde{\kvec}) 
&= \frac{1}{N^{d/2}} \sum_{\xvec \in \mathbb{K}_N^d} e^{- i \langle \xvec, \widetilde{\kvec} \rangle} \ g(\xvec),
\nonumber
\\
g(\xvec) 
&= \frac{1}{N^{d/2}} \sum_{\widetilde{\kvec} \in \widetilde{\mathbb{K}}_N^d} e^{i \langle \xvec, \widetilde{\kvec} \rangle} \ \widehat{g}(\widetilde{\kvec}),
\end{align*}
for $g : \mathbb{K}_N^d \longrightarrow \mathbb{C}^{2d}$ and $\widehat{g} : \widetilde{\mathbb{K}}_N^d \longrightarrow \mathbb{C}^{2d}$. In order to take a limit $N \to \infty$, we introduced the notation given in Eq. \eqref{kantan01}. We should note that as for the summation, we sometimes write ``$\kvec \in \mathbb{K}_N^d$" instead of ``$\widetilde{\kvec} \in \widetilde{\mathbb{K}}_N^d$". From the Fourier transform and Eq. \eqref{reunitaryop1}, we have
\begin{align*}
\widehat{\Psi}_{n+1}(\kvec)=\widehat{M}_A(\kvec)\widehat{\Psi}_n(\kvec),
\end{align*}
where $\Psi_n : T_N^d \longrightarrow \mathbb{C}^{2d}$ and $2d \times 2d$ matrix $\widehat{M}_A(\kvec)$ is determined by
\begin{align*}                          
\widehat{M}_A(\kvec)=\sum_{j=1}^{d} \Big( e^{2 \pi i k_j/N} P_{2j-1} A + e^{-2 \pi i k_j /N} P_{2j} A \Big). 
\end{align*}
By using notations in Eq. \eqref{kantan01}, we get 
\begin{align}                          
\widehat{M}_A(\widetilde{\kvec})=\sum_{j=1}^{d} \Big( e^{i \widetilde{k}_j} P_{2j-1} A + e^{-i \widetilde{k}_j} P_{2j} A \Big). 
\label{migiyoshi01}
\end{align}
Next we will consider the following eigenvalue problem for $2d N^d \times 2d N^d$ matrix $M_A$: 
\begin{align}
\lambda \Psi = M_A \Psi, 
\label{yoiko03}
\end{align}
where $\lambda \in \mathbb{C}$ is an eigenvalue and $\Psi (\in \mathbb{C}^{2d N^d})$ is the corresponding eigenvector. Noting that Eq. \eqref{yoiko03} is closely related to Eq. \eqref{reunitaryop1}, we see that Eq. \eqref{yoiko03} is rewritten as 
\begin{align}
\lambda \Psi (\xvec) = (M_A \Psi)(\xvec) = \sum_{j=1}^{d}\Big(P_{2j-1} A \Psi (\xvec+\evec_j)+P_{2j} A \Psi (\xvec-\evec_j)\Big),
\label{yoiko04}
\end{align} 
for any $\xvec \in \mathbb{K}_N^d$. From the Fourier transform and Eq. \eqref{yoiko04}, we obtain
\begin{align*}
\lambda \widehat{\Psi} (\kvec) = \widehat{M}_A (\kvec) \widehat{\Psi} (\kvec), 
\end{align*}
for any $\kvec \in \mathbb{K}_N^d$. Then the characteristic polynomials of $2d \times 2d$ matrix $\widehat{M}_A(\kvec)$ for fixed $\kvec (\in \mathbb{K}_N^d)$ is 
\begin{align}                          
\det \Big( \lambda I_{2d} - \widehat{M}_A (\kvec) \Big) = \prod_{j=1}^{2d} \Big( \lambda - \lambda_{j} (\kvec) \Big),
\label{yoiko06}
\end{align}
where $\lambda_{j} (\kvec)$ are eigenvalues of $\widehat{M}_A (\kvec)$. Similarly, the characteristic polynomials of $2d N^d \times 2d N^d$ matrix $\widehat{M}_A$ is 
\begin{align*}                          
\det \Big( \lambda I_{2d N^d} - \widehat{M}_A  \Big) = \prod_{j=1}^{2d} \prod_{\kvec \in \mathbb{K}_N^d} \Big( \lambda - \lambda_{j} (\kvec) \Big).
\end{align*}
Thus we have
\begin{align*}                          
\det \Big( \lambda I_{2d N^d} - M_A  \Big) 
&= \det \Big( \lambda I_{2d N^d} - \widehat{M}_A  \Big) = \prod_{j=1}^{2d} \prod_{\kvec \in \mathbb{K}_N^d} \Big( \lambda - \lambda_{j} (\kvec) \Big).
\end{align*}
Therefore, by taking $\lambda = 1/u$, we get the following key result.
\begin{align}                          
\det \Big( I_{2d N^d} - u M_A  \Big) 
= \det \Big( I_{2d N^d} - u \widehat{M}_A  \Big) = \prod_{j=1}^{2d} \prod_{\kvec \in \mathbb{K}_N^d} \Big( 1 - u \lambda_{j} (\kvec) \Big).
\label{keyresult01lemma}
\end{align}
We should note that for fixed $\kvec (\in \mathbb{K}_N^d)$, eigenvalues of  $2d \times 2d$ matrix $\widehat{M}_A (\kvec)$ are expressed as 
\begin{align*}
{\rm Spec} ( \widehat{M}_A (\kvec)) = \left\{  \lambda_{j} (\kvec) \ | \ j = 1, 2, \ldots, 2d \right\}. 
\end{align*}
Moreover, eigenvalues of $2d N^d \times 2d N^d$ matrix not only $\widehat{M}_A$ but also $M_A$ are expressed as 
\begin{align*}
{\rm Spec} ( \widehat{M}_A) = {\rm Spec} (M_A)= \left\{  \lambda_{j} (\kvec) \ | \ j = 1, 2, \ldots, 2d, \ \kvec \in \mathbb{K}_N^d \right\}. 
\end{align*}

By using notations in Eq. \eqref{kantan01} and Eq. \eqref{yoiko06}, we see that for fixed $\kvec (\in \mathbb{K}_N^d)$,  
\begin{align}                          
\det \Big( I_{2d} - u \widehat{M}_A (\widetilde{\kvec}) \Big) = \prod_{j=1}^{2d} \Big( 1 - u \lambda_{j} (\widetilde{\kvec}) \Big).
\label{yoiko08}
\end{align}
Furthermore, Eq. \eqref{migiyoshi01} gives the following important formula.
\begin{align*}                          
\det \Big( I_{2d} - u \widehat{M}_A (\widetilde{\kvec}) \Big) = \det \left(I_{2d} - u \times \sum_{j=1}^{d} \Big( e^{i \widetilde{k}_j} P_{2j-1} A + e^{-i \widetilde{k}_j} P_{2j} A \Big) \right). 
\end{align*}

In this setting, we define the {\em walk-type zeta function} by 
\begin{align}
\overline{\zeta} \left(A, T^d_N, u \right) = \det \Big( I_{2d N^d} - u M_A \Big)^{-1/N^d}.
\label{satosan01}
\end{align}
In general, for a $d_c \times d_c$ coin matrix $A$, we put  
\begin{align*}
\overline{\zeta} \left(A, T^d_N, u \right) = \det \Big(I_{d_c N^d} - u M_A \Big)^{-1/N^d}.
\end{align*}
We should remark that the walk-type zeta function becomes the generalized zeta function $\overline{\zeta} \left(T^d_N, u \right)$ in \cite{K1} for the Grover walk (F-type). So we write the walk-type zeta function with a coin matrix $A$ as $\overline{\zeta} \left(A, T^d_N, u \right)$. Moreover, our walk is defined on the ``site" $\xvec (\in T^d_N)$. On the other hand, the walk in \cite{K1} is defined on the ``arc" (i.e., oriented edge). However, both of the walks are the same for the torus case.

By Eqs. \eqref{keyresult01lemma}, \eqref{yoiko08} and \eqref{satosan01}, we get 
\begin{align*}
\overline{\zeta} \left(A, T^d_N, u \right) ^{-1}
=
\exp \left[ \frac{1}{N^d} \sum_{\widetilde{\kvec} \in \widetilde{\mathbb{K}}_N^d} \log \left\{ \det \Big( I_{2d} - u \widehat{M}_A (\widetilde{\kvec}) \Big) \right\} \right].
\end{align*}
Sometimes we write $\sum_{\kvec \in \mathbb{K}_N^d}$ instead of $\sum_{\widetilde{\kvec} \in \widetilde{\mathbb{K}}_N^d}$. Noting $\widetilde{k_j} = 2 \pi k_j/N \ (j=1,2, \ldots, d)$ and taking a limit as $N \to \infty$, we show
\begin{align*}
\lim_{N \to \infty} \overline{\zeta} \left(A, T^d_N, u \right) ^{-1}
=
\exp \left[ \int_{[0,2 \pi)^d} \log \left\{ \det \Big( I_{2d} - u \widehat{M}_A \left( \Theta^{(d)} \right) \Big) \right\} d \Theta^{(d)}_{unif} \right],
\end{align*}
if the limit exists. We should note that when we take a limit as $N \to \infty$, we assume that the limit exists throughout this paper. Here $\Theta^{(d)} = (\theta_1, \theta_2, \ldots, \theta_d) (\in [0, 2 \pi)^d)$ and $d \Theta^{(d)}_{unif}$ denotes the uniform measure on $[0, 2 \pi)^d$, that is,
\begin{align*}
d \Theta^{(d)}_{unif} = \frac{d \theta_1}{2 \pi } \cdots \frac{d \theta_d}{2 \pi }.
\end{align*}
Then the following result was obtained in \cite{K2}.
\begin{theorem}[Komatsu, Konno and Sato \cite{K2}]
\begin{align*}
\overline{\zeta} \left(A, T^d_N, u \right) ^{-1}
&= \exp \left[ \frac{1}{N^d} \sum_{\widetilde{\kvec} \in \widetilde{\mathbb{K}}_N^d} \log \left\{ \det \Big( F(\widetilde{\kvec}, u) \Big) \right\} \right],
\\
\lim_{N \to \infty} \overline{\zeta} \left(A, T^d_N, u \right) ^{-1}
&=
\exp \left[ \int_{[0,2 \pi)^d} \log \left\{ \det \Big( F \left( \Theta^{(d)}, u \right)  \Big) \right\} d \Theta^{(d)}_{unif} \right],
\end{align*}
where 
\begin{align*}
F \left( \wvec , u \right) = I_{2d} - u \widehat{M}_A (\wvec), 
\end{align*}
with $\wvec = (w_1, w_2, \ldots, w_d) \in \RM^d$.
\label{th001}
\end{theorem}

Furthermore, we define $C_r (A, T^d_N)$ by
\begin{align}
\overline{\zeta} \left(A, T^d_N, u \right) = \exp \left( \sum_{r=1}^{\infty} \frac{C_r (A, T^d_N)}{r} u^r \right).
\label{satosan03}
\end{align}
Sometimes we write $C_r (A, T^d_N)$ by $C_r$ for short. Combining Eq. \eqref{satosan01} with Eq. \eqref{satosan03} gives
\begin{align*}
\det \Big( I_{2d N^d} - u M_A \Big)^{-1/N^d} = \exp \left( \sum_{r=1}^{\infty} \frac{C_r}{r} u^r \right).
\end{align*}
Thus we get
\begin{align}
- \frac{1}{N^d} \log \left\{ \det \Big( I_{2d N^d} - u M_A \Big) \right\} =  \sum_{r=1}^{\infty} \frac{C_r}{r} u^r.
\label{satosan05}
\end{align}
It follows from Eq. \eqref{keyresult01lemma} that the left-hand of Eq. \eqref{satosan05} becomes 
\begin{align*}
- \frac{1}{N^d} \log \left\{ \det \Big( I_{2d N^d} - u \widehat{M}_A \Big) \right\} 
&= 
- \frac{1}{N^d} \sum_{j=1}^{2d} \sum_{\kvec \in \mathbb{K}_N^d} \log \left\{ 1 - u \lambda_{j} (\kvec) \right\} 
\\
&
= \frac{1}{N^d} \sum_{j=1}^{2d} \sum_{\kvec \in \mathbb{K}_N^d} \sum_{r=1}^{\infty} \frac{\left(\lambda_{j} (\kvec) \right)^r}{r} u^r. 
\end{align*}
By this and the right-hand of Eq. \eqref{satosan05}, we have
\begin{align}
C_r (A, T^d_N) = \frac{1}{N^d} \sum_{j=1}^{2d} \sum_{\kvec \in \mathbb{K}_N^d} \left(\lambda_{j} (\kvec) \right)^r 
= \frac{1}{N^d} \sum_{j=1}^{2d} \sum_{\widetilde{\kvec} \in \widetilde{\mathbb{K}}_N^d} \left(\lambda_{j} (\widetilde{\kvec}) \right)^r.
\label{satosan06t}
\end{align}
Noting $\widetilde{k_j} = 2 \pi k_j/N \ (j=1,2, \ldots, d)$ and taking a limit as $N \to \infty$, we get
\begin{align}
\lim_{N \to \infty} C_r (A, T^d_N) = \sum_{j=1}^{2d} \int_{[0,2 \pi)^d} \lambda_{j} \left( \Theta^{(d)} \right)^r d \Theta^{(d)}_{unif}.
\label{satosan06limit}
\end{align}
Let ${\rm Tr} (A)$ denote the trace of a square matrix $A$. Then by definition of ${\rm Tr}$ and Eqs. \eqref{satosan06t} and \eqref{satosan06limit}, the following result was shown in \cite{K2}.

\begin{theorem}[Komatsu, Konno and Sato \cite{K2}]
\begin{align*}
C_r (A, T^d_N) 
&
= \frac{1}{N^d} \sum_{\widetilde{\kvec} \in \widetilde{\mathbb{K}}_N^d} {\rm Tr} \left( \left( \widehat{M}_A (\widetilde{\kvec}) \right)^r \right),
\nonumber
\\
\lim_{N \to \infty} C_r (A, T^d_N) 
&
= \int_{[0,2 \pi)^d} {\rm Tr} \left( \left( \widehat{M}_A (\Theta^{(d)}) \right)^r \right) d \Theta^{(d)}_{unif}
= {\rm Tr} \left( \Phi_r ^{(\infty)} ({\bf 0}) \right).
\end{align*}
\label{satosan06prop}
\end{theorem}
An interesting point is that $\Phi_r ^{(\infty)} ({\bf 0})$ is the return ``matrix weight" at time $r$ for the walk on not $T_N^d$ but $\ZM^d$. We should remark that in general ${\rm Tr} ( \Phi_r ^{(\infty)} ({\bf 0}) )$ is not the same as the return probability at time $r$ for QW and CRW. 

From now on, we will present the result on only ``$\lim_{N \to \infty}$" for $\overline{\zeta} \left(A, T^d_N, u \right) ^{-1}$ and $C_r (A, T^d_N)$, since the corresponding expression for ``without $\lim_{N \to \infty}$" is the essentially same (see Theorems \ref{th001} and \ref{satosan06prop}, for example).

\section{One-Dimensional Three-State QW and CRW \label{sec03}} 
This section is devoted to the three-state QW (case (i)) and CRW (case (ii)) on the one-dimensional torus $T_N^1$. Remark that a detailed study on the two-state QW and CRW was given in \cite{K2}. 
\par
\
\par\noindent
(i) QW case. 
\par
We consider the following $3 \times 3$ coin matrix $A^{(m)}_{QW}$ (M-type) and $A^{(f)}_{QW}$ (F-type) introduced by Machida \cite{Machida2015}. 
\begin{align*}
A^{(m)}_{QW}=
\begin{bmatrix}
-\frac{1+\cos \eta }{2} & \frac{\sin \eta }{\sqrt{2}} & \frac{1-\cos \eta }{2}\\
\frac{\sin \eta }{\sqrt{2}} & \cos \eta  & \frac{\sin \eta }{\sqrt{2}}\\
\frac{1-\cos \eta }{2} & \frac{\sin \eta }{\sqrt{2}} & -\frac{1+\cos \eta }{2}
\end{bmatrix}, 
\quad 
A^{(f)}_{QW}=
\begin{bmatrix}
\frac{1-\cos \eta }{2} & \frac{\sin \eta }{\sqrt{2}} & -\frac{1+\cos \eta }{2}\\
\frac{\sin \eta }{\sqrt{2}} & \cos \eta  & \frac{\sin \eta }{\sqrt{2}}\\
-\frac{1+\cos \eta }{2} & \frac{\sin \eta }{\sqrt{2}} & \frac{1-\cos \eta }{2}
\end{bmatrix}, 
\end{align*}
for $\eta \in [0, 2\pi)$. If $\cos \eta=-1/3$, then the QW becomes the so-called Grover walk which is one of the most well-investigated model in the study of QWs. In this model, we take the projections $\{P_{-1}, P_0, P_1\}$ by 
\begin{align*}
P_{-1}=
\begin{bmatrix}
1 & 0 & 0\\
0 & 0 & 0\\
0 & 0 & 0
\end{bmatrix}, \quad
P_0=
\begin{bmatrix}
0 & 0 & 0\\
0 & 1 & 0\\
0 & 0 & 0
\end{bmatrix}, \quad
P_1=
\begin{bmatrix}
0 & 0 & 0\\
0 & 0 & 0\\
0 & 0 & 1
\end{bmatrix} .
\end{align*}
Then $3 \times 3$ matrix $\widehat{M}_{A^{(s)}_{QW}}(\widetilde{k})$ for $s \in \{m, f\}$ is defined by 
\begin{align*}
\widehat{M}_{A^{(s)}_{QW}}(\widetilde{k})
=e^{i\widetilde{k}} P_{-1}A^{(s)}_{QW}+P_0A^{(s)}_{QW}+e^{-i\widetilde{k}}P_1A^{(s)}_{QW}. 
\end{align*}
We see that the walker moves at each step one unit to the left with $P_{-1}A^{(s)}_{QW}$ or one unit to the right with $P_1 A^{(s)}_{QW}$ or stays with $P_0 A^{(s)}_{QW}$. Thus we get 
\begin{prop}
\begin{align}
\lim_{N \to \infty}\overline{\zeta}\left(A^{(s)}_{QW}, T_N^1, u\right)^{-1} 
=(1+(-1)^{\delta(s)}u)\exp\left[\int_0^{2\pi }\log\left\{F^{(s)}_{QW}( \theta, u)\right\}\frac{d\theta}{2\pi}\right], 
\label{kimarid13qwnf}
\end{align}
for $s \in \{m, f\}$, where $\delta(s)=1$ for $s=m$, $\delta(s)=0$ for $s=f$ and 
\begin{align*}
F^{(s)}_{QW}(\theta, u)
=1-\left[(-1)^{\delta(s)}+\cos\eta+\left\{(-1)^{\delta(s)}-\cos\eta \right\} \cos\theta \right]u+u^{2}. 
\end{align*}
\end{prop}
Remark that the leading factor $(1+(-1)^{\delta(s)} u)$ of the right-hand side of Eq. \eqref{kimarid13qwnf} corresponds to localization of the three-state Grover walk on $\ZM$ (see Machida \cite{Machida2015}, for example). Localization means that there exists an initial state such that limsup for time $n$ of the probability that the walker returns to the starting location at time $n$ is positive. Specially, in the case of $\cos \eta=-1/3$ (Grover walk), we have
\begin{cor}
\begin{align*}
F^{(s)}_{QW} \left( \theta, u \right) 
= 1 - \frac{(-1)^{\delta(s)} \cdot \ 2}{3}  \left( 1 + 2 \cos \theta + \delta(s) (1 - \cos \theta ) \right) u + u^2.
\end{align*}
\label{kimarid13qw}
\end{cor}
\noindent
This result corresponds to Corollary 8 in \cite{K2}. Moreover, we will compute $C_r(A^{(s)}_{QW}, T_N^1)$ for this QW. Then we see that the eigenvalues $\lambda_1, \lambda_2, \lambda_3$ of $\widehat{M}_{A^{(s)}_{QW}}(\widetilde{k})$ can be written as follows. 
\begin{align*}
\lambda_1=-(-1)^{\delta(s)},\quad \lambda_2=\frac{t+i\sqrt{4-t^2}}{2},\quad \lambda_3=\frac{t-i\sqrt{4-t^2}}{2},  
\end{align*}
where 
\begin{align*}
t = t(\widetilde{k})=(-1)^{\delta(s)}+\cos\eta+\left\{(-1)^{\delta(s)}-\cos\eta\right\} \cos\widetilde{k}. 
\end{align*}
Remark that unitarity of $\widehat{M}_{A^{(s)}_{QW}}(\widetilde{k})$ implies $|\lambda_1|=|\lambda_2|=|\lambda_3|=1$. From Theorem \ref{satosan06prop}, we get
\begin{prop}
\begin{align*}
\lim_{N \to \infty}C_r(A^{(s)}_{QW}, T_N^1)
=\int_{0}^{2\pi} G^{(s)}(\theta)\frac{d\theta}{2\pi}, 
\end{align*}
for $s \in \{m, f\}$, where 
\begin{align*}
G^{(s)}(\theta)
&=\left\{-(-1)^{\delta(s)}\right\}^r+\left(\frac{t+i\sqrt{4-t^2}}{2}\right)^r+\left(\frac{t-i\sqrt{4-t^2}}{2}\right)^r,
\\
t
&= t(\theta)=(-1)^{\delta(s)}+\cos\eta+\left\{(-1)^{\delta(s)}-\cos\eta\right\} \cos \theta. 
\end{align*}
\end{prop}

\noindent
(ii) CRW case. 

We consider the following $3 \times 3$ coin matrix $A^{(s)}_{CRW}$ given by using the Hadamard product of coin matrix $A^{(s)}_{QW}$ for $s \in \{m, f\}$. That is, 
\begin{align*}
A_{CRW}^{(m)}
=A_{QW}^{(m)} \odot A_{QW}^{(m)}
&=
\begin{bmatrix}
\frac{( 1+\cos \eta )^{2}}{4} & \frac{\sin^{2} \eta }{2} & \frac{( 1-\cos \eta )^{2}}{4}\\
\frac{\sin^{2} \eta }{2} & \cos^{2} \eta  & \frac{\sin^{2} \eta }{2}\\
\frac{( 1-\cos \eta )^{2}}{4} & \frac{\sin^{2} \eta }{2} & \frac{( 1+\cos \eta )^{2}}{4}
\end{bmatrix}, \\
A_{CRW}^{(f)}
=A_{QW}^{(f)} \odot A_{QW}^{(f)}
&=
\begin{bmatrix}
\frac{(1-\cos \eta)^{2}}{4} & \frac{\sin^{2} \eta }{2} & \frac{( 1+\cos \eta )^{2}}{4}\\
\frac{\sin^{2} \eta }{2} & \cos^{2} \eta  & \frac{\sin^{2} \eta }{2}\\
\frac{(1+\cos \eta)^{2}}{4} & \frac{\sin^{2} \eta }{2} & \frac{( 1-\cos \eta )^{2}}{4}
\end{bmatrix}, 
\end{align*}
where $\odot$ is the Hadamard product. For this model, similarly we obtain 
\begin{prop}
\begin{align*}
\lim_{N \to \infty}\overline{\zeta }\left( A^{(s)}_{CRW} ,T_{N}^{1} ,u\right)^{-1}
=\exp\left[\int _{0}^{2\pi }\log\left\{F^{(s)}_{CRW}(\theta, u)\right\}\frac{d\theta }{2\pi }\right], 
\end{align*}
for $s \in \{m, f\}$, where 
\begin{align*}
F^{(s)}_{CRW}(\theta, u)
&=1-\frac{u}{2}\left[\left\{1-(-1)^{\delta(s)}\cos\eta \right\}^2\cos \theta+2\cos^2\eta \right] \\
&-\frac{(-1)^{\delta(s)} \cdot u^2}{2}\left[\left\{1-(-1)^{\delta(s)}\cos\eta \right\}^2\left\{(-1)^{\delta(s)}+2\cos\eta \right\}\cos \theta+\cos\eta(1+\cos^2\eta) \right] \\
&-\frac{(-1)^{\delta(s)} \cdot u^3}{2}\cos\eta(1-3\cos^2\eta).
\end{align*}
Here $\delta(s)=1$ for $s=m$, $\delta(s)=0$ for $s=f$.
\end{prop}

\section{One-Dimensional Four-State QW and CRW \label{sec04}} 
In this section, we deal with the four-state QW (case (i)) and CRW (case (ii)) on the one-dimensional torus $T_N^1$.
\par
\
\par\noindent
(i) QW case. 
\par
We consider the following $4 \times 4$ coin matrix $A^{(m)}_{QW}$ (M-type) and $A^{(f)}_{QW}$ (F-type) introduced by Watabe et al. \cite{WatabeEtAl2008}.
\begin{align}
A^{(m)}_{QW}
&=
\begin{bmatrix}
p-1 & p & \sqrt{pq} & \sqrt{pq}\\
p & p-1 & \sqrt{pq} & \sqrt{pq}\\
\sqrt{pq} & \sqrt{pq} & q-1 & q\\
\sqrt{pq} & \sqrt{pq} & q & q-1
\end{bmatrix}, 
\label{akoreyoM}
\\
A^{(f)}_{QW}
&=
\begin{bmatrix}
\sqrt{pq} & \sqrt{pq} & q & q-1\\
\sqrt{pq} & \sqrt{pq} & q-1 & q\\
p & p-1 & \sqrt{pq} & \sqrt{pq}\\
p-1 & p & \sqrt{pq} & \sqrt{pq}
\end{bmatrix}
,
\label{akoreyoF}
\end{align}
where $p, q \in [0, 1]$ and $q=1-p$. If $p=q=1/2$, then the QW becomes the Grover walk. In this model, we take the projections $\{P_{-2}, P_{-1}, P_1, P_2\}$ by 
\begin{align*}
P_{-2}&=
\begin{bmatrix}
1 & 0 & 0 & 0\\
0 & 0 & 0 & 0\\
0 & 0 & 0 & 0\\
0 & 0 & 0 & 0
\end{bmatrix} ,& 
P_{-1}&=
\begin{bmatrix}
0 & 0 & 0 & 0\\
0 & 1 & 0 & 0\\
0 & 0 & 0 & 0\\
0 & 0 & 0 & 0
\end{bmatrix} ,& 
P_{1}&=
\begin{bmatrix}
0 & 0 & 0 & 0\\
0 & 0 & 0 & 0\\
0 & 0 & 1 & 0\\
0 & 0 & 0 & 0
\end{bmatrix} , & 
P_{2}&=
\begin{bmatrix}
0 & 0 & 0 & 0\\
0 & 0 & 0 & 0\\
0 & 0 & 0 & 0\\
0 & 0 & 0 & 1
\end{bmatrix} .
\end{align*}
For $s \in \{ m,f \}$, we define $4 \times 4$ matrix $\widehat{M}_{A^{(s)}_{QW}}(\widetilde{k})$ by
\begin{align*}
\widehat{M}_{A^{(s)}_{QW}}(\widetilde{k})
=e^{2i\widetilde{k}} P_{-2} A^{(s)}_{QW}+e^{i\widetilde{k}} P_{-1} A^{(s)}_{QW}+e^{-i\widetilde{k}} P_{1} A^{(s)}_{QW} + e^{-2i\widetilde{k}} P_{2} A^{(s)}_{QW}. 
\end{align*}
The walker moves at each step two unit to the left with $P_{-2} A^{(s)}_{QW}$ or one unit to the left with $P_{-1} A^{(s)}_{QW}$ or one unit to the right with $P_1 A^{(s)}_{QW}$ or two unit to the right with $P_2 A^{(s)}_{QW}$. Then we have 
\begin{prop}
\begin{align*}
\lim _{N \to \infty}\overline{\zeta}\left(A^{(m)}_{QW}, T_N^1, u\right)^{-1}
&=\exp\left[\int_0^{2\pi }\log\left\{F^{(m)}_{QW}(\theta, u)\right\}\frac{d\theta}{2\pi }\right], 
\\
\lim_{N \to \infty}\overline{\zeta }\left(A^{(f)}_{QW}, T_N^1, u\right)^{-1}
&=\left(1-u^{2}\right)\exp\left[\int_0^{2\pi}\log\left\{F^{(f)}_{QW}\left(\theta, u\right)\right\}\disf{d\theta}{2\pi}\right].
\end{align*}
Here
\begin{align*}
F^{(m)}_{QW}(\theta, u)
&=1+\left\{\cos\theta+\cos(2\theta)-2ip_\ast(\sin\theta+\sin(2\theta))\right\}u-4ip_\ast\sin(3\theta)u^2 
\\
&-\left\{\cos \theta+\cos(2\theta)+2ip_\ast(\sin\theta+\sin(2\theta))\right\}u^3-u^4, 
\\
F^{(f)}_{QW}(\theta, u)
&=1-\sqrt{1 - 4 p_{\ast}^2}\left(\cos \theta+\cos(2\theta)\right)u+u^2, 
\end{align*}
where $p_{\ast}=p-1/2 \ (\in [-1/2, 1/2])$. 
\label{hanzatsu01}
\end{prop}
As a special case, if $p_{\ast}=0$ ($p=1/2$), then the QW becomes the Grover walk. Then we obtain 
\begin{cor}
\begin{align*}
\lim _{N \to \infty}\overline{\zeta}\left(A^{(s)}_{QW}, T_N^1, u\right)^{-1}
=\left(1-u^2\right)\exp\left[\int_{0}^{2\pi}\log\left\{1  - (-1)^{\delta(s)} \left(\cos \theta +\cos(2\theta)\right) u+u^2\right\}\frac{d\theta}{2\pi}\right],
\end{align*}
where $\delta(s)=1$ for $s=m$, $\delta(s)=0$ for $s=f$. 
\end{cor}
We note that Inui and Konno \cite{IK} investigated our M-type model with $p_{\ast}=0$ case and showed localization occurs. In fact, Proposition \ref{hanzatsu01} implies that concerning the M-type, localization occurs for only $p_{\ast}=0$. In contrast to M-type model, localization occurs for F-type model with any $p_{\ast} \in [-1/2, 1/2]$.

From now on, we focus on F-type model, since $F^{(f)}_{QW}(\theta, u)$ is a quadratic polynomial with respect to $u$. Then the eigenvalues $\lambda$ of $M_{A^{(f)}_{QW}}(\widetilde{k})$ with $|\lambda|=1$ can be written as follows.
\begin{align*}
\lambda=\pm 1, \quad \alpha_q (\widetilde{k}) \pm i \sqrt{1 - \alpha_q (\widetilde{k})^2},
\end{align*}
where 
\begin{align*}
\alpha_q (\widetilde{k}) = \frac{1}{2} \sqrt{1 - 4 p_{\ast}^2} \ (\cos\widetilde{k}+\cos(2\widetilde{k})). 
\end{align*}
Then we obtain 
\begin{prop}
\begin{align*}
\lim_{N \to \infty}C_r(A^{(f)}_{QW}, T_N^1)
=\int_{0}^{2\pi} G^{(f)}_{QW}(\theta)\frac{d\theta}{2\pi}, 
\end{align*}
where 
\begin{align*}
G^{(f)}_{QW}(\theta)=1+(-1)^r+\left(\alpha_q (\theta) + i \sqrt{1 - \alpha_q (\theta)^2}\right)^r + \left(\alpha_q (\theta) - i \sqrt{1 - \alpha_q(\theta)^2}\right)^r. 
\end{align*}
\end{prop}

\noindent
(ii) CRW case. 

As in the case of the one-dimensional three-state CRW in the previous section, we treat the following $4 \times 4$ coin matrix $A^{(s)}_{CRW}$ given by using the Hadamard product of coin matrix $A^{(s)}_{QW}$ for $s \in \{m, f\}$ as follows.
\begin{align}
A_{CRW}^{(m)}
&=A_{QW}^{(m)} \odot A_{QW}^{(m)} =
\begin{bmatrix}
(p-1)^2 & p^2 & pq & pq \\
p^2 & (p-1)^2 & pq & pq \\
pq & pq & (q-1)^2 & q^2 \\
pq & pq & q^2 & (q-1)^2
\end{bmatrix}
\label{sousai01}, 
\\
A_{CRW}^{(f)}
&=A_{QW}^{(f)} \odot A_{QW}^{(f)} = 
\begin{bmatrix}
pq & pq & q^2 & (q-1)^2 \\
pq & pq & (q-1)^2 & q^2 \\
p^2 & (p-1)^2 & pq & pq \\
(p-1)^2 & p^2 & pq & pq
\end{bmatrix}
,
\label{sousai02}
\end{align}
where $p, q \in [0, 1]$ and $q=1-p$. If $p=q=1/2$, the CRW becomes a RW. For $s \in \{ m,f \}$, like QW, we define $4 \times 4$ matrix $\widehat{M}_{A^{(s)}_{CRW}}(\widetilde{k})$ by
\begin{align*}
\widehat{M}_{A^{(s)}_{CRW}}(\widetilde{k})
=e^{2i\widetilde{k}} P_{-2} A^{(s)}_{CRW}+e^{i\widetilde{k}} P_{-1} A^{(s)}_{CRW}+e^{-i\widetilde{k}} P_{1} A^{(s)}_{CRW} + e^{-2i\widetilde{k}} P_{2} A^{(s)}_{CRW}. 
\end{align*}
Thus we have 
\begin{prop}
\begin{align*}
\lim_{N \to \infty }\overline{\zeta }\left(A^{(m)}_{CRW}, T_N^1, u\right)^{-1}
&=\exp\left[\int_0^{2\pi}\log\left\{F^{(m)}_{CRW}(\theta, u)\right\}\disf{d\theta}{2\pi}\right], \\
\lim_{N \to \infty}\overline{\zeta }\left(A^{(f)}_{CRW}, T_N^1, u\right)^{-1}
&=(1+4p_\ast^2u^2)\exp\left[\int_0^{2\pi }\log\left\{F^{(f)}_{CRW}(\theta, u)\right\}\disf{d\theta}{2\pi}\right], 
\end{align*}
where $p_{\ast}=p-1/2$ and 
\begin{align*}
F^{(m)}_{CRW}(\theta, u)
&=1-\frac{1}{2}\left\{(1+4p_{\ast}^2)(\cos\theta+\cos(2\theta))-4ip_{\ast}(\sin\theta+\sin(2\theta))\right\}u
\\
&- 2ip_{\ast}(1+4p_{\ast}^2)\sin(3\theta)u^2 
\\
&+ 2p_{\ast}^2\left\{(1+4p_{\ast}^2)(\cos \theta+\cos(2\theta))+4ip_{\ast}(\sin \theta+\sin(2\theta))\right\}u^3-16p_{\ast}^4u^4,
\\
F^{(f)}_{CRW}(\theta, u)
&=1-\frac{1}{2}(1-4p_{\ast}^2)(\cos\theta+\cos(2\theta))u-4p_{\ast}^2u^2.
\end{align*}
\label{uhehe01}
\end{prop}
When $p_{\ast}=0 \ (p=1/2)$, our CRW becomes a four-state RW. In this model, Proposition \ref{uhehe01} implies 
\begin{equation*}
\lim_{N \to \infty}\overline{\zeta}\left(A^{(s)}_{CRW}, T_{N}^{1}\right)^{-1}
=\exp\left[\int_{0}^{2\pi}\left\{1-\frac{u}{2}(\cos\theta+\cos(2\theta))\right\}\frac{d\theta}{2\pi}\right],
\end{equation*}
for $s \in \{m, f\}$. Note that this result is equivalent to Corollary \ref{podo01} in Section \ref{sec05}.

From now on, we focus on F-type model, since $F^{(f)}_{CRW}(\theta, u)$ is a quadratic polynomial with respect to $u$. Then the eigenvalues $\lambda$ of $M_{A^{(f)}_{CRW}}(\widetilde{k})$ can be written as follows.
\begin{align*}
\lambda = \pm 2ip_{\ast}, \quad \alpha_c(\widetilde{k}) \pm \sqrt{\alpha_c(\widetilde{k})^2+4p_{\ast}^2}
\end{align*}
where 
\begin{align*}
\alpha_c(\widetilde{k})
=\frac{1}{4}(1-4p_{\ast})^2 \ (\cos\widetilde{k}+\cos(2\widetilde{k})). 
\end{align*}
Therefore we have 
\begin{prop}
\begin{align*}
\lim_{N \to \infty}C_r(A^{(f)}_{CRW}, T_N^1)
=\int_{0}^{2\pi} G^{(f)}_{CRW}(\theta)\frac{d\theta}{2\pi}, 
\end{align*}
where 
\begin{align*}
G^{(f)}_{CRW}(\theta)
=(2ip_{\ast})^r+(-2ip_{\ast})^r+\left(\alpha_c(\theta) + \sqrt{\alpha_c(\theta)^2+4p_{\ast}^2}\right)^r+\left(\alpha_c(\theta) - \sqrt{\alpha_c(\theta)^2+4p_{\ast}^2}\right)^r
\end{align*}
\end{prop}

\section{One-Dimensional Multi-State RW \label{sec05}}
In Sections \ref{sec03} and \ref{sec04}, we dealt with the three- and four-state QW and CRW on the one-dimensional torus $T_N^1$. When the number of the state is larger than five, the similar study becomes complicated. Therefore, we treat the RW with multi-state on $T_N^1$ in this section. The random walker considered here jumps at each step to location $x \in \{ -L, -(L-1), \ldots, L-1, L \}$ with probability $p_x$, where $L \in \ZM_{>}$. Here $\{ p_x \}$ satisfies $p_{x} \in [0, 1]$ and $\sum_{x=-L}^{L} p_{x} =1$. We should note that RW is a special case of CRW and its coin matrix $A_{RW}$ can be considered as real value $1$, that is, $A_{RW}=1$. So we get 
\begin{align*}
\widehat{M}_{A_{RW}}(\widetilde{k}) 
= \sum_{x=-L}^{L} e^{-i x \widetilde{k}} p_x.
\end{align*}
Then we obtain
\begin{prop}
\begin{align*}
\lim_{N \to \infty}\overline{\zeta}\left(A_{RW}, T_{N}^{1}, u\right)^{-1}
&=\exp\left\{\int_{0}^{2\pi }\log \left(1-u \left( \sum_{x=-L}^{L} e^{-i x \theta} p_x \right) \right) \frac{d\theta}{2\pi}\right\}, 
\\
\lim_{N \to \infty}C_r(A_{RW}, T_{N}^{1})
&=
\int_{0}^{2\pi} \left( \sum_{x=-L}^{L} e^{-i x \theta} p_x \right)^r \frac{d\theta}{2\pi}.
\end{align*}
\end{prop}

From now on, we consider the following special case: 
\begin{align*}
p_x = p_{\ast} \qquad (x \in \{-L, -(L-1), \ldots, L-1, L\} \backslash \{0\}). 
\end{align*}
That is, $p_x$ is the same as $p_{\ast}$ except the origin. So we have
\begin{align*}
p_0 + 2L \cdot p_{\ast} = 1 \quad (p_0, \ p_{\ast} \in [0, 1]). 
\end{align*}
Then we see that 
\begin{align*}
\widehat{M}_{A_{RW}}(\widetilde{k}) 
= p_{0} + 2 p_{\ast} \sum_{\ell=1}^{L}\cos(\ell \widetilde{k}). 
\end{align*}
Therefore we obtain 
\begin{prop}
\begin{align*}
\lim_{N \to \infty}\overline{\zeta }\left(A_{RW}, T_{N}^{1}, u \right)^{-1} 
&=\exp\left\{\int_{0}^{2\pi }\log \left(1-u \left( p_0 + 2 p_{\ast} \sum_{\ell=1}^{L} \cos (\ell \theta) \right) \right) \frac{d\theta}{2\pi}\right\}, 
\\
\lim_{N \to \infty}C_r(A_{RW}, T_{N}^{1})
&=
\int_{0}^{2\pi} \left( p_0 + 2 p_{\ast} \sum_{\ell=1}^{L} \cos (\ell \theta) \right)^r \frac{d\theta}{2\pi}.
\end{align*}
\label{akutagawa01}
\end{prop}

Then the following result is given by Proposition \ref{akutagawa01} for $L=1, \ p_0 = 0, \ p_{\ast}=1/2$ case.
\begin{cor}
\begin{align}
\lim_{N \to \infty}\overline{\zeta}\left(A_{RW}, T_N^1, u\right)^{-1}
&=\exp\left\{\int_0^{2\pi}\log(1-u\cos \theta)\frac{d\theta}{2\pi}\right\},
\label{akutagawa02}
\\
\lim_{N \to \infty}C_r(A_{RW}, T_{N}^{1})
&=
\int_{0}^{2\pi} \left(\cos \theta \right)^r \frac{d\theta}{2\pi}.
\nonumber
\end{align}
\end{cor}
We should note that this is equivalent to Corollary 7 in \cite{K2}. Moreover, for this case, in order to obtain an explicit form of the right-hand side of Eq. \eqref{akutagawa02}, we prepare the following lemma.
\begin{lemma}
\begin{align}
\log\left(\frac{1+\sqrt{1-x^{2}}}{2}\right)=-\sum_{n=1}^{\infty}\frac{1}{2n}\binom{2n}{n}\left(\frac{x^{2}}{4}\right)^{n}.
\label{akutagawa04}
\end{align}
\label{akutagawa05}
\end{lemma}
\par\noindent
{\bf Proof.} Let $f(x)$ be the left-hand side of Eq. \eqref{akutagawa04}. Differentiating $f(x)$ by $x$ gives 
\begin{align}
f'(x)=\frac{1}{x}\left\{1-\left(1-x^{2}\right)^{-\frac{1}{2}}\right\}. 
\label{relation01}
\end{align}
Here, the Taylor expansion of $\left(1-x^{2}\right)^{-1/2}$ implies the following formula. 
\begin{align}
\left(1-x^{2}\right)^{-\frac{1}{2}}
=\sum_{n=0}^{\infty}\binom{2n}{n}\left(\frac{1}{2}\right)^{2n}x^{2n}. 
\label{relation02}
\end{align}
Substituting Eq. \eqref{relation02} into Eq. \eqref{relation01} leads to 
\begin{align*}
f'(x)&=\frac{1}{x}\left\{1-\sum_{n=0}^{\infty }\binom{2n}{n}\left(\frac{1}{2}\right)^{2n} x^{2n}\right\}
=-\sum _{n=1}^{\infty }\binom{2n}{n}\left(\frac{1}{2}\right)^{2n}x^{2n-1}. 
\end{align*}
Therefore we have
\begin{align*}
f( x) & =\int _{0}^{x} f'( y) dy 
=-\sum_{n=1}^{\infty }\binom{2n}{n}\left(\frac{1}{2}\right)^{2n}\int _{0}^{x} y^{2n-1} dy 
=-\sum_{n=1}^{\infty }\frac{1}{2n}\binom{2n}{n}\left(\frac{1}{2}\right)^{2n} x^{2n}. 
\end{align*}
\hfill$\square$
\par
Now we have the following explicit form of the right-hand side of Eq. \eqref{akutagawa02}.
\begin{cor}
\begin{align*}
\lim_{N \to \infty}\overline{\zeta}\left(A_{RW}, T_N^1, u\right)^{-1}
= \frac{1+\sqrt{1-u^2}}{2}.
\end{align*}
\end{cor}
\par\noindent
{\bf Proof.} In order to compute the right-hand side of Eq. \eqref{akutagawa02}, we see that
\begin{align*}
\int_{0}^{2\pi}\log(1-u \cos\theta)\frac{d\theta}{2\pi}
&=-\frac{1}{\pi }\sum_{n=1}^{\infty}\frac{u^{2n}}{n}\int_{0}^{\frac{\pi}{2}}(\cos\theta)^{2n} d\theta \\
= - \sum_{n=1}^{\infty} \frac{1}{2n}\binom{2n}{n} \left(\frac{u^{2}}{4}\right)^{n}
&=\log\left(\frac{1+\sqrt{1-u^{2}}}{2}\right). 
\end{align*}
The last equality comes from Lemma \ref{akutagawa05}.
\hfill$\square$
\par
\
\par
Next the following result can be obtained by Proposition \ref{akutagawa01} for $L=2, \ p_0 = 0, \ p_{\ast}=1/4$ case.
\begin{cor}
\begin{align*}
\lim_{N \to \infty}\overline{\zeta}\left(A_{RW}, T_N^1, u\right)^{-1}
&=\exp\left[\int_0^{2\pi}\log \left\{1-\frac{u}{2}(\cos\theta+\cos(2\theta))\right\} \frac{d\theta}{2\pi}\right],
\\
\lim_{N \to \infty}C_r(A_{RW}, T_{N}^{1})
&=
\int_{0}^{2\pi} \left\{\frac{1}{2}(\cos\theta+\cos(2\theta))\right\}^r \frac{d\theta}{2\pi}.
\end{align*}
\label{podo01}
\end{cor}

Finally, we consider the case in which the walker moves at each step to each location with the same probability $p_0=p_{\ast}=1/(2L+1)$. In a similar fashion, 
we have
\begin{prop}
\begin{align*}
\lim_{N \to \infty}\overline{\zeta }\left(A_{RW}, T_{N}^{1}, u\right)^{-1}
&=\exp\left\{\int_{0}^{2\pi }\log \left(1-\frac{u}{2L+1}\frac{\cos((L+1)\theta)-\cos (L \theta)}{\cos \theta - 1} \right) \frac{d\theta}{2\pi}\right\}, 
\\
\lim_{N \to \infty} C_r(A_{RW}, T_N^1)
&=\int_0^{2\pi}\left\{\frac{1}{2L+1}\frac{\cos((L+1)\theta)-\cos (L\theta)}{\cos \theta-1}\right\}^r\frac{d\theta}{2\pi}. 
\end{align*}
\end{prop}

\section{Two-Dimensional Four-State QW and CRW \label{sec06}} 
In this section, we treat the four-state QW (case (i)) and CRW (case (ii)) on the two-dimensional torus $T_N^2$.
\par
\
\par\noindent
(i) QW case. 
\par
We consider the following $4 \times 4$ coin matrix $A^{(m)}_{QW}$ (M-type) and $A^{(f)}_{QW}$ (F-type). 
\begin{align}
A^{(m)}_{QW}
&=
\begin{bmatrix}
p-1 & p & \sqrt{pq} & \sqrt{pq}\\
p & p-1 & \sqrt{pq} & \sqrt{pq}\\
\sqrt{pq} & \sqrt{pq} & q-1 & q\\
\sqrt{pq} & \sqrt{pq} & q & q-1
\end{bmatrix}, 
\label{akoreyoM2} 
\\
A^{(f)}_{QW}
&=
\begin{bmatrix}
p & p-1 & \sqrt{pq} & \sqrt{pq}\\
p-1 & p & \sqrt{pq} & \sqrt{pq}\\
\sqrt{pq} & \sqrt{pq} & q & q-1\\
\sqrt{pq} & \sqrt{pq} & q-1 & q
\end{bmatrix}
= \left(I_{2} \otimes \sigma \right) A^{(m)}_{QW},
\label{akoreyoF2}
\end{align}
where $p, q \in [0, 1]$ and $q=1-p$. If $p=q=1/2$, then the QW becomes the Grover walk. Note that Eq. \eqref{akoreyoM2} equals Eq. \eqref{akoreyoM}, however, Eq. \eqref{akoreyoF2} (given by Eq. \eqref{goho}) does not equal Eq. \eqref{akoreyoF}. In this model, we take the projections $\{ P_{1}, P_{2}, P_3, P_3 \}$ by 
\begin{align*}
P_{1}&=
\begin{bmatrix}
1 & 0 & 0 & 0\\
0 & 0 & 0 & 0\\
0 & 0 & 0 & 0\\
0 & 0 & 0 & 0
\end{bmatrix} ,& 
P_{2}&=
\begin{bmatrix}
0 & 0 & 0 & 0\\
0 & 1 & 0 & 0\\
0 & 0 & 0 & 0\\
0 & 0 & 0 & 0
\end{bmatrix} ,& 
P_{3}&=
\begin{bmatrix}
0 & 0 & 0 & 0\\
0 & 0 & 0 & 0\\
0 & 0 & 1 & 0\\
0 & 0 & 0 & 0
\end{bmatrix} , & 
P_{4}&=
\begin{bmatrix}
0 & 0 & 0 & 0\\
0 & 0 & 0 & 0\\
0 & 0 & 0 & 0\\
0 & 0 & 0 & 1
\end{bmatrix} .
\end{align*}
For $s \in \{ m,f \}$, we define $4 \times 4$ matrix $\widehat{M}_{A^{(s)}_{QW}}(\widetilde{k}_{1}, \widetilde{k}_{2})$ as follows.
\begin{align*}
\widehat{M}_{A^{(s)}_{QW}}(\widetilde{k}_{1}, \widetilde{k}_{2})
=e^{i\widetilde{k}_{1}} P_{1} A^{(s)}_{QW} + e^{-i\widetilde{k}_{1}} P_{2} A^{(s)}_{QW} + e^{i\widetilde{k}_{2}} P_{3} A^{(s)}_{QW} + e^{-i\widetilde{k}_{2}} P_{4} A^{(s)}_{QW}. 
\end{align*}
The walker moves at each step one unit to the left with $P_1 A^{(s)}_{QW}$ or one unit to the right with $P_2 A^{(s)}_{QW}$ or one unit to the down with $P_3 A^{(s)}_{QW}$ or one unit to the up with $P_4 A^{(s)}_{QW}$. 
Thus we obtain 
\begin{prop}
\begin{align*}
\lim _{N \to \infty}\overline{\zeta }\left(A^{(s)}_{QW}, T_{N}^{2}, u\right)^{-1}
=\left(1-u^{2}\right)\exp\left[\int_{0}^{2\pi}\int_{0}^{2\pi}\log\left\{F^{(s)}(\theta_1, \theta_2, u)\right\}\frac{d\theta_1 }{2\pi}\frac{d\theta_2}{2\pi}\right], 
\end{align*}
for $s \in \{m, f\}$. Here $p_{\ast}=p-1/2$ and 
\begin{align*}
&F^{(s)}_{QW}(\theta_1, \theta_2, u)
\\
& \quad =1-(-1)^{\delta(s)} \left[\left\{1+(-1)^{\delta(s)} \cdot 2p_{\ast}\right\} \cos \theta_{1}+\left\{1-(-1)^{\delta(s)} \cdot 2p_{\ast}\right\}\cos \theta_{2}\right]u+u^2, 
\end{align*}
where $\delta(s)=1$ for $s=m$, $\delta(s)=0$ for $s=f$. 
\label{manrei01}
\end{prop}

Specially, Proposition \ref{manrei01} for $p=1/2$ (Grover walk) gives
\begin{cor}
\begin{align*}
F^{(s)}_{QW}(\theta_1, \theta_2, u)
=1 - (-1)^{\delta(s)} (\cos \theta_{1}+\cos \theta_{2})u+u^2,
\end{align*}
for $s \in \{m, f\}$. Here $\delta(s)=1$ for $s=m$, $\delta(s)=0$ for $s=f$. 
\end{cor}
\noindent
This result is equivalent to Corollary 11 in \cite{K2}. Moreover, for $s \in \{m, f\}$, the matrix $M_{A^{(s)}_{QW}}(\widetilde{k})$ has eigenvalues $\lambda$ with $|\lambda|=1$ as follows.
\begin{align*}
\lambda=\pm1, \quad \beta_q^{(s)} (\widetilde{k}_1, \widetilde{k}_2) \pm i \sqrt{ 1 - \beta_q^{(s)} (\widetilde{k}_1, \widetilde{k}_2)^2}, 
\end{align*}
where $p_{\ast}=p-1/2$ and 
\begin{align*}
\beta_q^{(s)} (\widetilde{k}_1, \widetilde{k}_2) = \left(\frac{1}{2}-\delta(s)+p_{\ast}\right) \cos\widetilde{k}_1+ \left(\frac{1}{2}-\delta(s)-p_{\ast}\right)\cos\widetilde{k}_2. 
\end{align*}
Then we have
\begin{prop}
\begin{align*}
\lim_{N \to \infty}C_r(A^{(s)}_{QW}, T_N^2)
=\int_{0}^{2\pi}\int_{0}^{2\pi} G^{(s)}_{QW}(\theta_1, \theta_2)\frac{d\theta_1}{2\pi}\frac{d\theta_2}{2\pi}, 
\end{align*}
for $s \in \{m, f\}$, where 
\begin{align*}
G^{(s)}_{QW} (\theta_1, \theta_2)
&=1+(-1)^r+\left(\beta_q^{(s)} (\theta_1, \theta_2) + i \sqrt{1 - \beta_q^{(s)} (\theta_1, \theta_2)^2} \right)^r 
\\
&+\left( \beta_q^{(s)} (\theta_1, \theta_2) - i \sqrt{1 - \beta_q^{(s)} (\theta_1, \theta_2)^2} \right)^r,
\\
\beta_q^{(s)} (\theta_1, \theta_2)
&= \left(\frac{1}{2}-\delta(s)+p_{\ast}\right) \cos \theta_1+ \left(\frac{1}{2}-\delta(s)-p_{\ast}\right)\cos \theta_2. 
\end{align*}
\end{prop}

\noindent
(ii) CRW case. 

As in the case of the one-dimensional four-state CRW in Section \ref{sec04}, we treat the following $4 \times 4$ coin matrix $A^{(s)}_{CRW}$ given by using the Hadamard product of coin matrix $A^{(s)}_{QW}$ for $s \in \{m, f\}$ as follows.
\begin{align}
A_{CRW}^{(m)}
&=A_{QW}^{(m)} \odot A_{QW}^{(m)} =
\begin{bmatrix}
(p-1)^2 & p^2 & pq & pq \\
p^2 & (p-1)^2 & pq & pq \\
pq & pq & (q-1)^2 & q^2 \\
pq & pq & q^2 & (q-1)^2
\end{bmatrix}
\label{sousai01two}, 
\\
A_{CRW}^{(f)}
&=A_{QW}^{(f)} \odot A_{QW}^{(f)} = 
\begin{bmatrix}
p^2 & (p-1)^2 & pq & pq \\
(p-1)^2 & p^2 & pq & pq \\
pq & pq & q^2 & (q-1)^2 \\
pq & pq & (q-1)^2 & q^2 
\end{bmatrix}
= \left(I_{2} \otimes \sigma \right) A^{(m)}_{CRW}
,
\label{sousai02two}
\end{align}
where $p, q \in [0, 1]$ and $q=1-p$. If $p=q=1/2$, the CRW becomes a RW. Note that Eq. \eqref{sousai01two} is equal to Eq. \eqref{sousai01}, however, Eq. \eqref{sousai02two} (given by Eq. \eqref{goho}) is not equal to Eq. \eqref{sousai02}. For $s \in \{ m,f \}$, like QW, we define $4 \times 4$ matrix $\widehat{M}_{A^{(s)}_{CRW}}(\widetilde{k}_{1}, \widetilde{k}_{2})$ by
\begin{align*}
\widehat{M}_{A^{(s)}_{CRW}}(\widetilde{k}_{1}, \widetilde{k}_{2})
=e^{i\widetilde{k}_{1}} P_{1} A^{(s)}_{CRW} + e^{-i\widetilde{k}_{1}} P_{2} A^{(s)}_{CRW} + e^{i\widetilde{k}_{2}} P_{3} A^{(s)}_{CRW} + e^{-i\widetilde{k}_{2}} P_{4} A^{(s)}_{CRW}. 
\end{align*}
Then we get
\begin{prop}
\begin{align*}
\lim _{N\rightarrow \infty }\overline{\zeta }\left( A^{(s)}_{CRW} ,T_{N}^{2} ,u\right)^{-1}
=(1-4p_{\ast}^2u^2)\exp\left[\int _{0}^{2\pi }\int _{0}^{2\pi }\log\left\{F^{( s)}_{CRW} (\theta _{1} ,\theta _{2} ,u)\right\}\frac{d\theta _{1}}{2\pi }\frac{d\theta _{2}}{2\pi }\right], 
\end{align*}
for $s \in \{m, f\}$. Here 
\begin{align*}
&F^{(s)}_{CRW} (\theta_{1}, \theta_{2}, u)
\\
& \quad =1-\frac{1}{2}\left[\left\{(-1)^{\delta(s)}+2p_{\ast}\right\}^{2}\cos \theta_{1} +\left\{(-1)^{\delta(s)}-2p_{\ast}\right\}^{2}\cos \theta_{2}\right] u+4p_{\ast }^2u^2,
\end{align*}
where $p_{\ast}=p-1/2$. 
\end{prop}

Furthermore, for $s \in \{m, f\}$, the eigenvalues $\lambda$ of $M_{A^{(s)}_{CRW}}(\widetilde{k})$ can be written as follows.
\begin{align*}
\lambda=\pm2p_{\ast}, \quad \beta_c^{(s)} (\widetilde{k}_1, \widetilde{k}_2) \pm \sqrt{\beta_c^{(s)} (\widetilde{k}_1, \widetilde{k}_2)^2-4p_{\ast}^2}, 
\end{align*}
where 
\begin{align*}
\beta_c^{(s)} (\widetilde{k}_1, \widetilde{k}_2) = \left(\frac{1}{2}-\delta(s)+p_{\ast}\right)^2 \cos \widetilde{k}_1+ \left(\frac{1}{2}-\delta(s)-p_{\ast}\right)^2 \cos \widetilde{k}_2. 
\end{align*}
Then we have 
\begin{prop}
\begin{align*}
\lim_{N \to \infty}C_r(A^{(s)}_{CRW}, T_N^2)
=\int_{0}^{2\pi}\int_{0}^{2\pi} G^{(s)}_{CRW}(\theta_1, \theta_2)\frac{d\theta_1}{2\pi}\frac{d\theta_2}{2\pi}, 
\end{align*}
for $s \in \{m, f\}$, where 
\begin{align*}
G^{(s)}_{CRW}(\theta_1, \theta_2)
&=(2p_{\ast})^r+(-2p_{\ast})^r + \left( \beta_c^{(s)} (\theta_1, \theta_2) + \sqrt{\beta_c^{(s)} (\theta_1, \theta_2)^2 -4p_{\ast}^2} \right)^r 
\\
& + \left( \beta_c^{(s)} (\theta_1, \theta_2)-\sqrt{\beta_c^{(s)} (\theta_1, \theta_2)^2-4p_{\ast}^2} \right)^r,
\\
\beta_c^{(s)} (\theta_1, \theta_2) 
&= \left(\frac{1}{2}-\delta(s)+p_{\ast}\right)^2 \cos \theta_1+ \left(\frac{1}{2}-\delta(s)-p_{\ast}\right)^2 \cos \theta_2. 
\end{align*}
\end{prop}

\section{Generalized Grover Matrix \label{sec07}} 
In this section, we introduce a new class of models determined by the {\em generalized Grover matrix} $U(a)$ with parameter $a \in [0,1]$ which connects the positive-support of the Grover matrix ($a=0$) and the Grover matrix ($a=1$). In fact, $d_c \times d_c$ generalized Grover matrix $U(a)=[U(a)_{ij}]$ is defined by
\begin{align*}
U(a)_{ij}= \left( \frac{2}{d_c} - 1 \right) a + 1 - \delta_{ij}
=
\begin{cases}
\left( \frac{2}{d_c} - 1 \right) a & (i=j), \\
\left( \frac{2}{d_c} - 1 \right) a + 1 & (i \neq j),
\end{cases}
\end{align*}
for $a \in [0, 1]$, where $\delta_{ij} = 1 \ (i=j), = 0 \ (i \not= j)$. We should remark that if $d_c =2$, then $U(a)$ does not depend on $a$. That is, 
\begin{align*}
U(a)=
\begin{bmatrix}
0 & 1 \\
1 & 0
\end{bmatrix}
\qquad (a \in [0, 1]).
\end{align*}
Thus $U(a)$ is unitary for any $a \in [0,1]$ if $d_c =2$. On the other hand, we see that ``$U(a)$ is unitary" if and only if ``$a=1$" for $d_c \ge 3$.

From now on, we compute zeta functions for the models determined by $U(a)$. First, as in a similar fashion of Section \ref{sec03}, we deal with the three-state model on the one-dimensional torus $T_N^1$ defined by the following $3 \times 3$ coin matrix $U^{(m)}(a)$ (M-type) and $U^{(f)}(a)$ (F-type). 
\begin{align*}
U^{(m)} (a) =
\begin{bmatrix}
-\frac{a}{3} & -\frac{a}{3}+1 & -\frac{a}{3}+1 \\
-\frac{a}{3}+1 & -\frac{a}{3} & -\frac{a}{3}+1 \\
-\frac{a}{3}+1 & -\frac{a}{3}+1 & -\frac{a}{3} 
\end{bmatrix},
\quad 
U^{(f)}(a)=
\begin{bmatrix}
-\frac{a}{3}+1 & -\frac{a}{3}+1 & -\frac{a}{3} \\
-\frac{a}{3}+1 & -\frac{a}{3} & -\frac{a}{3}+1 \\
-\frac{a}{3} & -\frac{a}{3}+1 & -\frac{a}{3}+1 
\end{bmatrix}. 
\end{align*}
By a similar argument in Section \ref{sec03}, we have 
\begin{prop}
\begin{align*}
\lim_{N \to \infty }\overline{\zeta }\left(U^{(m)}(a), T_N^1, u\right)^{-1}
&=\exp\left[\int_0^{2\pi}\log\left\{F^{(m)}(\theta, u, a)\right\}\disf{d\theta}{2\pi}\right], \\
\lim_{N \to \infty }\overline{\zeta }\left(U^{(f)}(a), T_N^1, u\right)^{-1}
&=(1+u)\exp\left[\int_0^{2\pi}\log\left\{F^{(f)}(\theta, u, a)\right\}\disf{d\theta}{2\pi}\right], 
\end{align*}
where 
\begin{align*}
F^{(m)}(\theta, u, a) 
&=1+\frac{a}{3}(1+2\cos \theta)u+\frac{2a-3}{3}(1+2\cos \theta)u^2+(a-2)u^3, \\
F^{(f)}(\theta, u, a) 
&=1+\frac{a-3}{3}(1+2\cos \theta)u-(a-2)u^2.
\end{align*}
\end{prop}


Moreover, similarly in Section \ref{sec04}, we treat the four-state model on the one-dimensional torus $T_N^1$ defined by the following $4 \times 4$ coin matrix $U^{(m)}(a)$ (M-type) and $U^{(f)}(a)$ (F-type). 
\begin{align}
U^{(m)}(a) &=
\begin{bmatrix}
-\frac{a}{2} & -\frac{a}{2}+1 & -\frac{a}{2}+1 & -\frac{a}{2}+1 \\
-\frac{a}{2}+1 & -\frac{a}{2} & -\frac{a}{2}+1 & -\frac{a}{2}+1 \\
-\frac{a}{2}+1 & -\frac{a}{2}+1 & -\frac{a}{2} & -\frac{a}{2}+1 \\
-\frac{a}{2}+1 & -\frac{a}{2}+1 & -\frac{a}{2}+1 & -\frac{a}{2} \\
\end{bmatrix}, 
\label{koreyoM}
\\
U^{(f)} (a) &=
\begin{bmatrix}
-\frac{a}{2}+1 & -\frac{a}{2}+1 & -\frac{a}{2}+1 & -\frac{a}{2} \\
-\frac{a}{2}+1 & -\frac{a}{2}+1 & -\frac{a}{2} & -\frac{a}{2}+1 \\
-\frac{a}{2}+1 & -\frac{a}{2} & -\frac{a}{2}+1 & -\frac{a}{2}+1 \\
-\frac{a}{2} & -\frac{a}{2}+1 & -\frac{a}{2}+1 & -\frac{a}{2}+1 \\
\end{bmatrix}. 
\label{koreyoF}
\end{align}
Then we get the following result. 
\begin{prop}
\begin{align*}
\lim_{N \to \infty }\overline{\zeta }\left(U^{(m)}(a), T_N^1, u\right)^{-1}
&=\exp\left[\int_0^{2\pi}\log\left\{F^{(m)}(\theta, u, a)\right\}\disf{d\theta}{2\pi}\right], \\
\lim_{N \to \infty }\overline{\zeta }\left(U^{(f)}(a), T_N^1, u\right)^{-1}
&=(1-u^2)\exp\left[\int_0^{2\pi}\log\left\{F^{(f)}(\theta, u, a)\right\}\disf{d\theta}{2\pi}\right], 
\end{align*}
where 
\begin{align*}
F^{(m)}(\theta, u, a) 
&=1+a(\cos \theta+\cos2\theta)u+2(a-1)(1+\cos \theta+\cos3\theta)u^2 \\
&+(3a-4)(\cos\theta+\cos2\theta)u^3+(2a-3)u^4, \\
F^{(f)}(\theta, u, a) 
&=1+(a-2)(\cos\theta+\cos 2\theta)u-(2a-3)u^2. 
\end{align*}
\end{prop}

Finally, as in a similar fashion of Section \ref{sec06}, we consider the four-state model on the two-dimensional torus $T_N^2$ defined by the following $4 \times 4$ coin matrix $U^{(m)}(a)$ (M-type) and $U^{(f)}(a)$ (F-type). 
\begin{align}
U^{(m)}(a) &=
\begin{bmatrix}
-\frac{a}{2} & -\frac{a}{2}+1 & -\frac{a}{2}+1 & -\frac{a}{2}+1 \\
-\frac{a}{2}+1 & -\frac{a}{2} & -\frac{a}{2}+1 & -\frac{a}{2}+1 \\
-\frac{a}{2}+1 & -\frac{a}{2}+1 & -\frac{a}{2} & -\frac{a}{2}+1 \\
-\frac{a}{2}+1 & -\frac{a}{2}+1 & -\frac{a}{2}+1 & -\frac{a}{2} 
\end{bmatrix}, 
\label{koreyoM2}
\\
U^{(f)} (a) &=
\begin{bmatrix}
-\frac{a}{2}+1 & -\frac{a}{2} & -\frac{a}{2}+1 & -\frac{a}{2}+1 \\
-\frac{a}{2} & -\frac{a}{2}+1 & -\frac{a}{2}+1 & -\frac{a}{2}+1 \\
-\frac{a}{2}+1 & -\frac{a}{2}+1 & -\frac{a}{2}+1 & -\frac{a}{2} \\
-\frac{a}{2}+1 & -\frac{a}{2}+1 & -\frac{a}{2} & -\frac{a}{2}+1 
\end{bmatrix}
= \left(I_{2} \otimes \sigma \right) U^{(m)}(a). 
\label{koreyoF2}
\end{align}
Remark that Eq. \eqref{koreyoM2} is equal to Eq. \eqref{koreyoM}, however, Eq. \eqref{koreyoF2} (given by Eq. \eqref{goho}) is not equal to Eq. \eqref{koreyoF}. Then we show 
\begin{prop}
\begin{align*}
\lim_{N \to \infty }\overline{\zeta }\left(U^{(m)}(a), T_N^2, u\right)^{-1}
&=\exp\left[\int_0^{2\pi}\int_0^{2\pi}\log\left\{F^{(m)}(\theta_1, \theta_2, u, a)\right\}\disf{d\theta_1}{2\pi}\frac{d\theta_2}{2\pi}\right], 
\\
\lim_{N \to \infty }\overline{\zeta }\left(U^{(f)}(a), T_N^2, u\right)^{-1}
&=(1-u^2) 
\exp\left[\int_0^{2\pi}\int_0^{2\pi}\log\left\{F^{(f)}(\theta_1, \theta_2, u,a)\right\}\disf{d\theta_1}{2\pi}\frac{d\theta_2}{2\pi}\right], 
\end{align*}
where
\begin{align*}
F^{(m)}(\theta_1, \theta_2, u, a) 
&=1+a(\cos \theta_1+\cos \theta_2)u+2(a-1)(1+2\cos \theta_1\cos \theta_2)u^2 \\
&+(3a-4)(\cos \theta_1+\cos \theta_2)u^3+(2a-3)u^4, \\
F^{(f)}(\theta_1, \theta_2, u, a) 
&=1+(a-2)(\cos \theta_1+\cos \theta_2)u-(2a-3)u^2. 
\end{align*}
\label{sugoiyo01}
\end{prop}

As for the M-type model, if $a= 0$ and $a=1$, then we have the following result for the positive-support of Grover matrix and the Grover matrix, respectively. 
\begin{cor}
\begin{align*}
\lim_{N \to \infty }\overline{\zeta }\left(U^{(m)}(0), T_N^2, u\right)^{-1}
&=\exp\left[\int_0^{2\pi}\int_0^{2\pi}\log\left\{F^{(m)}(\theta_1, \theta_2, u, 0)\right\}\disf{d\theta_1}{2\pi}\frac{d\theta_2}{2\pi}\right], 
\\
\lim_{N \to \infty }\overline{\zeta }\left(U^{(m)}(1), T_N^2, u\right)^{-1}
&=(1-u^2) \exp\left[\int_0^{2\pi}\int_0^{2\pi}\log\left\{F^{(m)}(\theta_1, \theta_2, u, 1)\right\}\disf{d\theta_1}{2\pi}\frac{d\theta_2}{2\pi}\right], 
\end{align*}
where 
\begin{align*}
F^{(m)}(\theta_1, \theta_2, u, 0) 
&=1-2(1+2\cos \theta_1\cos \theta_2)u^2-4(\cos \theta_1+\cos \theta_2)u^3-3u^4, 
\\
F^{(m)}(\theta_1, \theta_2, u, 1) 
&=1 + (\cos \theta_1 + \cos \theta_2)u + u^2.
\end{align*}
\end{cor}
We should remark that the results for $a=0$ and $a=1$ correspond to Corollaries 13 and 11 in \cite{K2}, respectively.


As for the F-type model, if $a= 0$ and $a=1$, then we have the following result for the positive-support of Grover matrix and the Grover matrix, respectively. 
\begin{cor}
\begin{align*}
\lim_{N \to \infty }\overline{\zeta }\left(U^{(f)}(0), T_N^2, u\right)^{-1}
&=(1-u^2) \exp\left[\int_0^{2\pi}\int_0^{2\pi}\log\left\{F^{(f)}(\theta_1, \theta_2, u,0)\right\}\disf{d\theta_1}{2\pi}\frac{d\theta_2}{2\pi}\right], 
\\
\lim_{N \to \infty }\overline{\zeta }\left(U^{(f)}(1), T_N^2, u\right)^{-1}
&=(1-u^2) \exp\left[\int_0^{2\pi}\int_0^{2\pi}\log\left\{F^{(f)}(\theta_1, \theta_2, u, 1)\right\}\disf{d\theta_1}{2\pi}\frac{d\theta_2}{2\pi}\right], 
\end{align*}
where 
\begin{align*}
F^{(f)}(\theta_1, \theta_2, u, 0) 
&=1-2(\cos \theta_1+\cos \theta_2)u+3u^2, 
\\
F^{(f)}(\theta_1, \theta_2, u, 1) 
&=1-(\cos \theta_1+\cos \theta_2)u+u^2. 
\end{align*}
\label{itti01}
\end{cor}
We should note that the result for $a=0$ corresponds to Corollary 13 in \cite{K2} and Eq. (10) in Clair \cite{Clair}, and the result for $a=1$ corresponds to Corollary 11 in \cite{K2}.

\section{Relation Between Grover/Zeta and Walk/Zeta Correspondences \label{sec08}}
In this section, we consider a relation between Grover/Zeta Correspondence in \cite{K1} and Walk/Zeta Correspondence in \cite{K2}. 

Let $G=(V(G),E(G))$ be a simple connected graph with the set $V(G)$ of vertices and the set $E(G)$ of unoriented edges $uv$ joining two vertices $u$ and $v$. Furthermore, let $n=|V(G)|$ and $m=|E(G)|$ be the number of vertices and edges of $G$, respectively. For $uv \in E(G)$, an arc $(u,v)$ is the oriented edge from $u$ to $v$. Let $D_G$ be the symmetric digraph corresponding to $G$. Set $D(G)= \{ (u,v),(v,u) \mid uv \in E(G) \} $. For $e=(u,v) \in D(G)$, set $u=o(e)$ and $v=t(e)$. Furthermore, let $e^{-1}=(v,u)$ be the {\em inverse} of $e=(u,v)$. For $v \in V(G)$, the {\em degree} $\deg {}_G v = \deg v = d_v $ of $v$ is the number of vertices adjacent to $v$ in $G$. Let $G$ be a connected graph with $n$ vertices $v_1, \ldots ,v_n $. The {\em adjacency matrix} ${\bf A}= {\bf A} (G)= [a_{ij}]$ is the square matrix such that $a_{ij} =1$ if $v_i$ and $v_j$ are adjacent, and $a_{ij} =0$ otherwise. If $ \deg {}_G v=k$ (constant) for each $v \in V(G)$, then $G$ is called {\em $k$-regular}. Moreover, the $n \times n$ matrix $P_{n} = {\bf P} (G) = [ P_{uv} ]_{u,v \in V(G)}$ is given as follows: 
\begin{align*}
P_{uv} =\left\{
\begin{array}{ll}
1/( \deg {}_G u)  & \mbox{if $(u,v) \in D(G)$, } \\
0 & \mbox{otherwise.}
\end{array}
\right.
\end{align*} 
Note that the matrix ${\bf P} (G)$ is the transition probability matrix of the simple random walk on $G$. If $G$ is a $(q+1)$-regular graph, then we have ${\bf P} (G)= \frac{1}{q+1} {\bf A} (G)$.

In this setting, Konno and Sato \cite{KS} presented the following result which is called the {\em Konno-Sato theorem} here.

\begin{theorem}[Konno and Sato \cite{KS}]
Let $G$ be a simple connected $(q+1)$-regular graph with $n$ vertices and $m$ edges for $q \in \ZM_{>}$. Then we have 
\begin{align*}  
\det \left( I_{2m} - u \ U^{(f)}(0) \right)
&=(1-u^2)^{m-n} \det \left( (1+qu^2) I_{n} -(q+1)u P_{n} \right),
\\  
\det \left( I_{2m} - u \ U^{(f)}(1)  \right)
&=(1-u^2)^{m-n} \det \left( (1+u^2) I_{n} -2u P_{n} \right).
\end{align*}  
\label{ksthm01}
\end{theorem}

Moreover, for the generalized Grover matrix $U(a)$ with $a \in [0,1]$ bridging the gap between the positive-support of the Grover matrix ($a=0$) and the Grover matrix ($a=1$), we can extend the Konno-Sato theorem (Theorem \ref{ksthm01}) as follows.
\begin{theorem}
Let $G$ be a simple connected $(q+1)$-regular graph with $n$ vertices and $m$ edges for $q \in \ZM_{>}$. Then we have
\begin{align*}  
&\det \left( I_{2m} - u \ U^{(f)}(a) \right)
\\
& \quad = (1-u^2)^{m-n} 
\det \left[ \left\{ 1+ \left( q +(1-q) a \right) u^2 \right\} I_{n} - 
\left\{ 1+q + (1-q) a \right\} u \ P_{n} \right],
\end{align*}
for $a \in [0,1]$.  
\label{ksthm}
\end{theorem}
\par\noindent
{\bf Proof.} The proof is the almost same as that of Theorem 4.1 in \cite{KS}, so we omit the details. The essential point is that we change $B_{ef}, \  w_{uv} , \ d_{uv}$ in the following way.
\begin{align*} 
B_{ef}
&= \frac{2}{d_{o(f)}} \quad \to \quad \left( \frac{2}{d_{o(f)}} - 1 \right) a + 1 = \frac{(1-a)d_{o(f)} + 2a}{d_{o(f)}} \qquad (e, f \in D(G)),
\\
w_{uv} 
&= \frac{2}{d_v} \quad \to \quad \left( \frac{2}{d_v} - 1 \right) a + 1 = \frac{(1-a)d_v + 2a}{d_v} \qquad (u, v \in V(G)),
\\
d_{uv} 
&= 2 \quad \to \quad (1-a)d_v + 2a \qquad (u, v \in V(G)).
\end{align*}
\hfill$\square$
\par
\
\par
We should remark that our walk with F-Type in Walk/Zeta Correspondence is defined on the ``site" $\xvec (\in T^d_N)$, on the other hand, the walk in Grover/Zeta Correspondence of \cite{K1} is defined on the ``arc" (i.e., oriented edge). However, both of the walks are the same for the torus case. This relation holds for the model given by a generalized Grover matrix $U(a)$ with $a \in [0,1]$.

Finally, we focus on the model determined by the $2d \times 2d$ generalized Grover matrix $U(a)$ with $a \in [0,1]$ on the $d$-dimensional torus $T^d_N$ which is $2d$-regular with $n=N^d$ vertices and $m=d N^d$ edges. By using a similar method in ``Section 6 Torus case" in \cite{K1}, from Theorem \ref{ksthm}, we obtain the following more general result. Indeed, $a=0$ case corresponds to Corollary 2 in \cite{K1} (Grover/Zeta Correspondence) and Corollary 16 in \cite{K2} (Walk/Zeta Correspondence), and $a=1$ case corresponds to Corollary 1 in \cite{K1} (Grover/Zeta Correspondence) and Corollary 14 in \cite{K2} (Walk/Zeta Correspondence). 
\begin{cor}
\begin{align*}
\overline{\zeta}\left(U^{(f)}(a), T_N^d, u \right)^{-1}
&=(1-u^2)^{d-1} 
\exp\left[\disf{1}{N^d}\sum_{\widetilde{\kvec} \in \widetilde{\mathbb{K}}_N^d }\log\left\{F^{(f)}\left(\widetilde{\kvec}, u, a \right)\right\}\right], \\
\lim_{N \to \infty }\overline{\zeta }\left(U^{(f)}(a), T_N^d, u\right)^{-1}
&=(1-u^2)^{d-1} 
\exp\left[\int_{[0, 2 \pi)^d} \log\left\{F^{(f)}(\Theta^{(d)}, u,a)\right\} d \Theta^{(d)}_{unif} \right],
\end{align*}
where
\begin{align*}
F^{(f)}(\wvec, u, a) 
=1 - \frac{2(d + (1-d) a)}{d} \left( \sum_{j=1}^d \cos w_j \right) \cdot u + \left\{ 2d-1 + 2 (1-d) a \right\} u^2, 
\end{align*}
for $a \in [0,1].$ Here $\widetilde{\kvec} =(\widetilde{k}_1, \ldots, \widetilde{k}_d) \in \widetilde{\mathbb{K}}_N^d, \ \Theta^{(d)} = (\theta_1, \theta_2, \ldots, \theta_d) \in [0, 2 \pi)^d, \ \wvec = (w_1, w_2, \ldots, w_d) \in \RM^d$, and $d \Theta^{(d)}_{unif}$ denotes the uniform measure on $[0, 2 \pi)^d$, that is,
\begin{align*}
d \Theta^{(d)}_{unif} = \frac{d \theta_1}{2 \pi } \cdots \frac{d \theta_d}{2 \pi }.
\end{align*}
\label{koredayo01}
\end{cor}
\noindent
Note that this result comes from an approach based on Grover/Zeta Correspondence. In particular, Corollary \ref{koredayo01} for $d=2$ case is equivalent to Proposition \ref{sugoiyo01} for F-type model, which is given by another approach based on Walk/Zeta Correspondence.

\section{Summary \label{sec09}} 

Following the previous paper on Walk/Zeta Correspondence by the first author and his coworkers \cite{K2}, we calculated the zeta function for various extended classes of QWs and CRWs on the torus determined by a more general coin matrix via the Fourier analysis. In Section \ref{sec02}, we reviewed Walk/Zeta Correspondence on the torus studied in \cite{K2}. In Sections \ref{sec03} and \ref{sec04}, we investigated the three- and four-state QW and CRW on the one-dimensional torus. Section \ref{sec05} treated the multi-state RW on the one-dimensional torus. Section \ref{sec06} dealt with the four-state QW and CRW on the two-dimensional torus. In Section \ref{sec07}, we introduced a new class of models determined by the generalized Grover matrix, which connects the Grover matrix and the positive-support of the Grover matrix. Section \ref{sec08} presented a generalized version of the Konno-Sato theorem for the new class. As a corollary, we computed the zeta function for the generalized Grover matrix on the $d$-dimensional torus. Moreover, we mentioned the relation between Grover/Zeta and Walk/Zeta Correspondences. To extend our class of models determined by the generalized Grover matrix $U(a)$ to a more general class would be one of the interesting problems \cite{K7}.




\end{document}